\documentclass[aps,pre,preprint,superscriptaddress]{revtex4-1}
\usepackage{bm}
\usepackage[dvips]{graphicx}

\makeatletter
\def\@cite#1{(#1)}
\makeatother
\begin{document}

\title{Inflow process of pedestrians to a confined space}
%\titlerunning{Short form of title} % if too long for running head

\author{Takahiro Ezaki}\email{ezaki@nii.ac.jp}
\affiliation{Japan Society for the Promotion of Science, 8 Ichibancho, Kojimachi, Chiyoda-ku, Tokyo 102-8472, Japan}
\affiliation{Research Center for Advanced Science and 
Technology, The University of Tokyo, 4-6-1 Komaba, Meguro-ku, 
Tokyo 153-8904, Japan}

\author{Kazumichi Ohtsuka}
\affiliation{Research Center for Advanced Science and 
Technology, The University of Tokyo, 4-6-1 Komaba, Meguro-ku, 
Tokyo 153-8904, Japan}
\author{Mohcine Chraibi}
\affiliation{J\"{u}lich Supercomputing Centre, 
Forschungszentrum J\"{u}lich GmbH, 52425 J\"{u}lich, Germany}

\author{Maik Boltes}
\affiliation{J\"{u}lich Supercomputing Centre, 
Forschungszentrum J\"{u}lich GmbH, 52425 J\"{u}lich, Germany}

\author{Daichi Yanagisawa}
\affiliation{Research Center for Advanced Science and 
Technology, The University of Tokyo, 4-6-1 Komaba, Meguro-ku, 
Tokyo 153-8904, Japan}

\author{Armin Seyfried}
\affiliation{J\"{u}lich Supercomputing Centre, 
Forschungszentrum J\"{u}lich GmbH, 52425 J\"{u}lich, Germany}
\affiliation{Department of Computer Simulation for Fire Safety and 
Pedestrian Traffic, Bergische Universit\"{a}t Wuppertal, 
42285 Wuppertal, Germany}

\author{Andreas Schadschneider}
\affiliation{Institut f\"{u}r Theoretische Physik, 
Universit\"{a}t zu K\"{o}ln, 50937 K\"{o}ln, Germany}

\author{Katsuhiro Nishinari}
\affiliation{Research Center for Advanced Science and 
Technology, The University of Tokyo, 4-6-1 Komaba, Meguro-ku, 
Tokyo 153-8904, Japan}

\begin{abstract}
To better design safe and comfortable urban spaces, understanding the nature of human crowd movement is important. 
However, precise interactions among pedestrians are difficult to measure in the presence of 
their complex decision-making processes and many related factors. 
While extensive studies on pedestrian flow through bottlenecks and corridors have been conducted, 
the dominant mode of interaction in these scenarios may not be relevant in different scenarios. 
Here, we attempt to decipher the factors that affect human reactions to other individuals from a different perspective. 
We conducted experiments employing the inflow process in which pedestrians successively enter a confined area (like an elevator) and look for a temporary position. 
In this process, pedestrians have a wider range of options regarding their motion than in the classical scenarios; therefore, 
other factors might become relevant.
The preference of location is visualized by pedestrian density profiles obtained from recorded pedestrian trajectories. 
Non-trivial patterns of space acquisition, e.g., an apparent preference for positions near corners, 
were observed. This indicates the relevance of psychological and anticipative factors beyond the private sphere,
which have not been deeply discussed so far in the literature on pedestrian dynamics.
From the results, four major factors, which we call flow avoidance, distance cost, angle cost, and boundary preference, were suggested.
We confirmed that a description of decision-making based on these factors can give a rise to realistic preference patterns, using a simple mathematical model. 
Our findings provide new perspectives and a baseline for considering the optimization of design and safety in crowded public areas and public transport carriers.
\end{abstract}

\keywords{pedestrian dynamics \and inflow process \and personal space}
\maketitle

%\tableofcontents

%%%%%%%%%%%%%%%%%%%%%%%%%%%%%%%%%%%%%%%%%%%%%%%%%%%%%%%%%%%%%%%%%%%
\section{Introduction}
\label{sec:intro}

Human crowding in public areas is still a significant issue
in social and engineering science \cite{Altman1975,Hillier1993}.  To 
design comfortable and safe urban spaces, it is important to
understand the nature of the interactions among pedestrians and their
consequences.  However, due to the lack of such knowledge, many
inefficient or even disastrous situations are still found, e.g.,
serious overcrowding in transportation systems in urban areas and crowding during disasters, leading to a stampede.  Meanwhile, relatively recent attempts have
partly succeeded in modeling the collective phenomena of pedestrians
\cite{Schadschneider2009,Helbing2001,Helbing2005b,Schadschneider2010}, including spontaneous lane
formation in a bidirectional flow \cite{Helbing1995,Burstedde2001,Kretz2006,Nowak2012},
crowd turbulence in overcrowded areas \cite{Helbing2007,Yu2007}, and formation of
stripe patterns in intersecting flows \cite{Helbing2005b,Ando1988}.  In
addition to experimental
\cite{Hoogendoorn2005,Kretz2006,Seyfried2009,Jelic2012,Yanagisawa2012a,Moussaid2012}
and empirical \cite{Henderson1971,Weidmann1993,Moussaid2010} studies,
several types of mathematical models that
facilitate simulations and analyses of central phenomena have been developed (see \cite{Schadschneider2009,Helbing2001,Helbing2005b,Schadschneider2010} for reviews).
Such approaches have also contributed to the understanding of how walking people recognize and react to
their environment. 

In this paper we study the inflow process of pedestrians that was
recently proposed as a model process for understanding the passenger
entrance behavior to elevators, buses, trains, etc. \cite{Ezaki2012c,Ezaki2013a}.  The
inflow process is defined by successive entry of
pedestrians
% \footnote{In order to avoid confusion, we use the term ``pedestrians'' to refer to standing people as well.} 
into a confined area and their subsequent dwelling.  
While entering, each pedestrian performs a complex decision-making process perceiving the current situation in the room (e.g.,
the distribution of persons), anticipating the behavior of
subsequently entering persons and planning to the exit from the room.
These stimuli and knowledge are evaluated for the decision, taking into account social
conventions. Among pedestrians, there is competition to optimize the final location with respect to easy and
fast exit under the limitation for avoiding overcrowding to secure comfort.
Classical studies of pedestrian dynamics focused on motion in corridors
and bottlenecks to investigate flow capacities or jamming at high
densities \cite{Schadschneider2009,Helbing2001,Helbing2005b,Schadschneider2010}.  In such scenarios, the decision process is negligible,
because pedestrians are not given the freedom of choice (e.g., changing the direction or
their locations) because of predefined destinations in the setup of the
experiment or restrictions on the motion at high densities.  In the
inflow process, no predefined destinations are given.  
Our goal is to reveal the modes of interactions among pedestrians that become visible through the decisions.
The distribution of pedestrians' positions could result from the interplay among several factors. 
A dominant mechanism could be based on the
concept of personal space, the area individuals maintain around
themselves into which others cannot intrude without causing discomfort
\cite{Hall1962,Little1965,Sommer1959}.  
Note that the effect of personal space has been integrated into computational models in various contexts in pedestrian dynamics \cite{Was2006,Was2010,Ezaki2012c,Manenti2012}. 

To understand such human behavior, further experimental and empirical studies are needed.
Recent behavioral experiments on the inflow process \cite{Ezaki2013a,Liu2016} succeeded in revealing that 
pedestrians prefer the areas near the boundaries, which had been predicted using a cellular automata model \cite{Ezaki2012c}.
In addition, Liu {\it et al}. \cite{Liu2015,Liu2016} extended the experiments described in Ref. \cite{Ezaki2013a} by varying the pedestrian number and the motivation for future exiting.

The number of pedestrians, i.e., the final pedestrian density in the area is considered to be a significant factor in pedestrian behavior because the main mechanism of the inflow process is expected to be strongly related to the avoidance reactions against other pedestrians.
If the final density is unknown or changes during the experimental trials, 
subjects might decide their final positions based on an incorrect understanding of the future situations.
To easily control this factor, we restrict ourselves to considering a fixed pedestrian number and area size in this study. Instead, we consider the geometry of the room as a variable in our experiment. 
As minor changes in the geometry of the area, we focus on (i) differences in entrance position and (ii) presence of an obstacle.
A previous study \cite{Ezaki2012c} suggested that the entrance position strongly affects the pedestrian distribution pattern.
In fact, these experimental settings allowed us to obtain robust average behavior and analyze the underlying decision-making.
Using pedestrian trajectory detection \cite{Boltes2013} and density
estimation \cite{Zhang2011} techniques developed in the context of
pedestrian dynamics, we attempt to visualize the behavior and the
  resulting decisions of pedestrians in such situations, thereby
  providing an insight into the anticipation and evaluation of future
  situations regarding their personal space.

In this study, we arranged a basic experiment of the inflow process (Fig.~\ref{set}). 
Test subjects were asked to enter a confined area, and temporarily stay there as described in Sec. \ref{experiment}.
On the basis of the obtained results, we report their interpretations (Sec. \ref{results}). 
Details of techniques and analyses used are summarized in Appendix. 
Finally we discuss the implications and limitations of the results in Sec. \ref{discuss}.

\section{Experimental Design}\label{experiment}

%%%%%%%%%%%%%%%%%%%%%%%%%%%%%%%%%%%%%%%%%%%%%%%%%%%%%%%%%%%%%%%%%%%%%%%%

\begin{figure}[t]
\centering
\includegraphics[width=150mm]{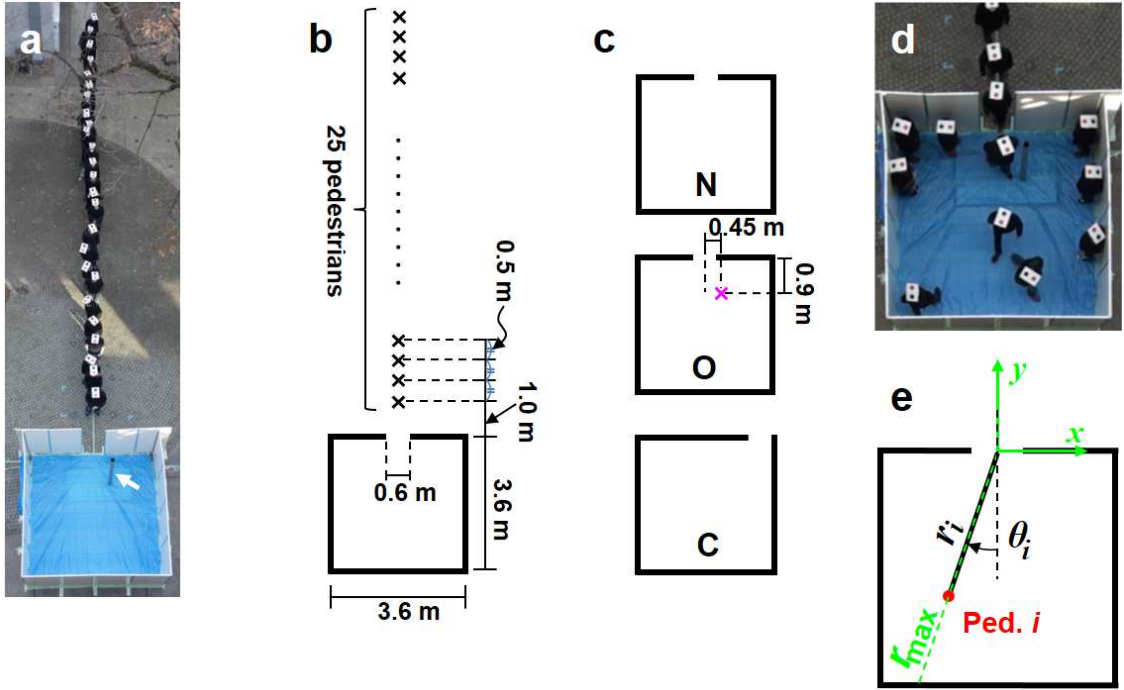}
\caption{(a) Snapshot of the experimental setup 
  (obstacle scenario). The pole indicated by a white arrow is the
  obstacle.  (b) Initial conditions and dimensions of the setup. (c)
  Three different scenarios (normal, N; obstacle, O; corner, C). The
  obstacle is shown by a magenta cross.  (d) Snapshot of the
  experiment. Pedestrians wore on their heads a piece of cardboard with two dots, by which head directions, as well as positions,
  were detected.  (e) Definitions of the $xy$-coordinate, the polar
  coordinate, ($r_i,\theta_i$), and the distance to the boundary,
  $r_{\rm{max}}$. }
\label{set}
\end{figure}

Controlled experiments have been conducted in the Research Center for
Advanced Science and Technology (RCAST) at The University of Tokyo,
Japan. A total of 25 male participants were recruited from
students of The University of Tokyo, who were paid for participation. 
The participants were strangers to each other.
The experiment was conducted for three different room structures (see Fig.~\ref{set}c): (i) With the
entrance in the middle without obstacles (``normal scenario''; N);  (ii) With the entrance in the middle with an
  obstacle in front (``obstacle scenario''; O); (iii) With the
entrance near a corner but without an obstacle (``corner
scenario''; C). The obstacle scenario ``O'' is included because it has
been suggested for evacuation processes of pedestrians that an
obstacle facilitates faster exiting by impeding clogging
\cite{Helbing2005b,Yanagisawa2009}. However, its (potentially, adverse)
effect on the inflow process is still unknown.  In addition, the
obstacle models the effects of hand-rail poles in buses or trains,
which is important within the transportation engineering context.

In each trial, 25 participants
were asked to enter an area (3.6 m$\times$ 3.6 m) set up on the
ground, which was enclosed within walls with a height of 2.0 m (Fig.~\ref{set}a
and Fig.~\ref{set}b). A PVC pipe ($\phi$0.11 m; 2.0 m high) was used as an
obstacle.  The initial positions of the pedestrians were marked on the
ground to regulate the inflow speed, and their entrance order was
randomized after each round.  They were instructed not to hurry but
behave as they would when boarding an elevator. The area was covered with a
blue sheet to cover the support structure of the walls and the obstacle, which were
sufficiently thin. The width of the entrance was 0.6 m. 
This value was chosen not to let the participants see inside the area before entrance in order to control the information about 
the configuration of pedestrians inside.

The experimental area was recorded with a digital camera (SONY
HDR-SR7; recording format 1080i60) mounted at a window on the 4th
floor (21 m high) of a neighboring building.  The participants were
equipped with caps and black shirts for video tracking, and their
trajectories were collected using tracking software (PeTrack
\cite{Boltes2013}).  We asked the participants to remain in the area for a
while after the final participant stopped walking and then asked them to
exit the room through the entrance for the next trial.  We repeated nine
trials for each scenario.  Prior to the experiment, we conducted a
few test trials using the normal scenario to let the participants
know the size of the area, speed of walking, etc., to prevent them
from being bewildered in an unfamiliar environment. 
Owing to these test trials, we observed no apparent difference in their behavioral patterns across experimental trials.

\section{Results}\label{results}

\subsection{Qualitative analysis of collective characteristics}

%\com{("Emergent cooperation in minimization of discomfort"
%= interpretation)}

\begin{figure}[t]
\centering
\includegraphics[width=150mm]{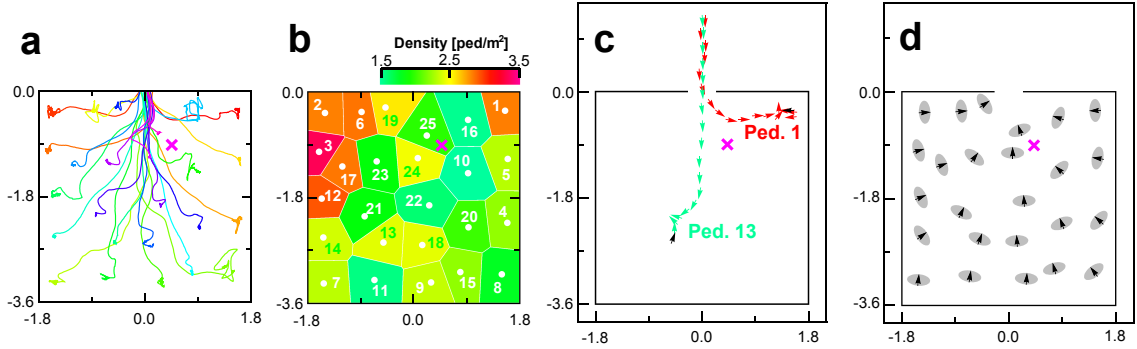}
\caption{Detection of pedestrian motion in a single trial 
  (obstacle scenario). (a) Trajectories of pedestrians. (b) Final
  distribution of pedestrians with Voronoi cells. Each number
  represents the entrance order. The local density is defined as the
  reciprocal of each Voronoi area (see equation (\ref{ld}) in Appendix).  (c) Sample trajectories with head
  directions. Black arrows correspond to the final state. (d) Head
  directions for a final state. For visualization of the space usage,
  referential body areas (not obtained from experimental data) are
  shown with ellipses of 0.4 m width and 0.2 m thickness.}
\label{figs}
\end{figure}

Fig.~\ref{figs} shows the pedestrian movement and final positions during one experiment.  
Locations near the boundaries were occupied first (Fig.~\ref{set}d, Fig.~\ref{figs}a and Fig.~\ref{figs}b). After stopping at their location, they remained standing there except for small fluctuations (Fig.~\ref{figs}a; see also supplementary video).  Interestingly, in this occupation order,
pedestrians were not required to and did not choose to pass between two closely located
pedestrians, minimizing the invasion of personal space.  The final distribution of pedestrians was uniform to a certain extent, with slight deviations in Voronoi areas (Fig.~\ref{figs}b). For larger distortions of the uniform state, pedestrians would fill up the gaps to gain more space, which in turn reduced the inhomogeneity. Rather homogeneous final state was thus achieved through a process of
self-organization.  The very homogeneous final distribution was reached
without significant changes to the first choice of destination, which
is a indication of good anticipation.

Just before reaching their final positions, pedestrians turn toward the
entrance (exit) where changes in the situation owing to successively
entering pedestrians are expected (Fig.~\ref{figs}c; see also supplementary video). 
This turning explains the features
near the end of the trajectories (Fig.~\ref{figs}a). 
The resulting
vector field of head directions is aligned toward the entrance
(Fig.~\ref{figs}d). This coordinated alignment is
  universally observed in all three experimental scenarios (N, O, and
  C). 

This behavior could be interpreted as the process of reducing the discomfort from not being able to see the next pedestrians entering the area. This potentially allows avoidance motion in case of conflicts. 
As explained, the personal space of pedestrians is fixed from the boundary of the area, and therefore, the direction for the boundary contains less uncertainty,
whereas stimuli requiring a reaction are mainly expected from the direction of the entrance. 
Furthermore, pedestrians who entered the space first and fixed their positions are
face-to-face with incoming pedestrians, potentially leading to
discomfort at eye contact. It is known that more personal space is
necessary when a person is confronted by others \cite{Argyle1965}. Therefore,
the alignment of head directions supports a more efficient use of space, with respective of the level of comfort.

Note that across the three scenarios, we observed no significant difference in the time required for 25 pedestrians to enter the area. 

\subsection{Qualitative analysis of individual decisions}

\begin{figure}[t]
\centering
\includegraphics[width=100mm]{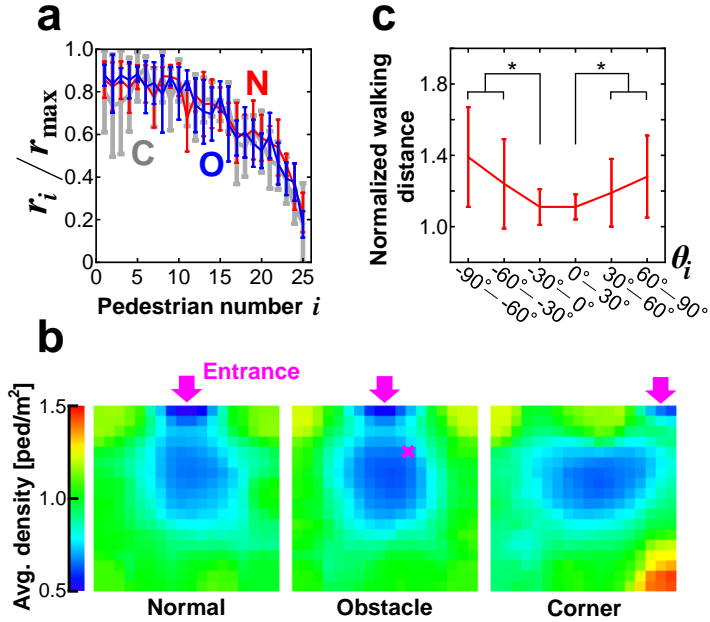}
\caption{(a) Normalized distance from the entrance to the final position of pedestrians. Each error bar represents the standard deviation ($N=9$ trials).  
  (b) Time and trial averaged density profiles using Voronoi
  diagrams.  Between scenarios~N and O, no significant difference was
  observed, whereas scenario~C displayed a distinctive pattern.  (c) Mean normalized walking distance with the standard deviation for the normal scenario.
The normalized walking distance was obtained
  by dividing the walking distance by the shortest distance between
  the entrance and the pedestrian's final position (see Eq. (\ref{nw})). As the turning angle increased, the walking
  distance was more extended.  The same feature was also observed for
  the other two scenarios.  Each asterisk represents statistical
  significance assessed by the Welch's $t$ test ($P<0.05$).}
\label{ir}
\end{figure}

Next, we study pedestrians' choice of location. By introducing polar
coordinates as shown in Fig.~\ref{set}e, the location of pedestrian
$i$ is expressed as ($r_i, \theta_i$). For the corner scenario, the
origin of these coordinates is in the corner where the entrance is located,
  i.e., at $(x,y) =(1.8,0)$. We also define $r_{\rm{max}}(\theta_i)$
for each direction $\theta_i$, which denotes the distance between the
entrance and the wall in that direction. To evaluate the positions of pedestrians, we use the normalized
distance $r_i/r_{\rm{max}}\in [0,1]$ for a fair comparison between different directions.
Fig.~\ref{ir}a shows a decrease of this ratio as the number of pedestrians in the room increases. 
This clearly illustrates that
pedestrians fill the area from the boundary walls to the entrance. 
This could be interpreted as anticipative behavior because the positions near boundaries are generally farther away from the entrance (exit). The normalized distance for N and O cases
approximately follow the same curve.  Meanwhile, in scenario~C, the
deviation of the normalized distance for the first four pedestrians is
larger than for the other two scenarios, which reflects the
inhomogeneity in the pedestrian distribution, which we will discuss next.

Fig.~\ref{ir}b displays the degree of space occupancy (the average pedestrian density integrated with respect to time for each trial; see also Appendix \ref{density profile}). 
High occupancy values indicate positions that tend to be already occupied at an
early stage in each trial. In scenarios N and O with the entrance in the middle, boundaries are uniformly preferred, except for the corners near the entrance, which are more attractive.  
In contrast, in scenario C, an \textit{asymmetric} distribution with respect to a diagonal line starting at the entrance is observed. In addition to the top-left and bottom-right corners, the top-middle boundary attracts pedestrians, showing characteristics distinct from those of the scenarios with a middle entrance.  This
preference for the top-middle boundary contributes to the large deviations in the normalized distance
(Fig.~\ref{ir}a). These qualitative differences in distributions
due to the entrance position indicate the motivation
of pedestrians in the process, which could be also viewed as a strong indicator for anticipation. Pedestrians entering in later
stages have fewer options for finding positions, usually
limited to the middle of the area.  For a better understanding of the
underlying mechanisms, it is therefore important to focus on the choice
of position of the pedestrians entering in the early stages of the
experiment.

\subsection{Interpretation and theoretical description}\label{interpretations}

As we have seen, the decision for which position to take is made based on anticipation. 
We discuss the factors that could affect the location choice of pedestrians to interpret the results based on the following assumptions.
First, the decision is made by considering an expected ideal configuration based on the knowledge of the total number of persons.
In this article, pedestrians anticipate an ideal uniform distribution, upon which they base their evaluation of each location.  
Second, a trade-off between finding a desirable location and minimizing costs
to reach there is considered, such that an option that reconciles these factors is realized. 

\begin{figure}[t]
\centering
\includegraphics[width=120mm]{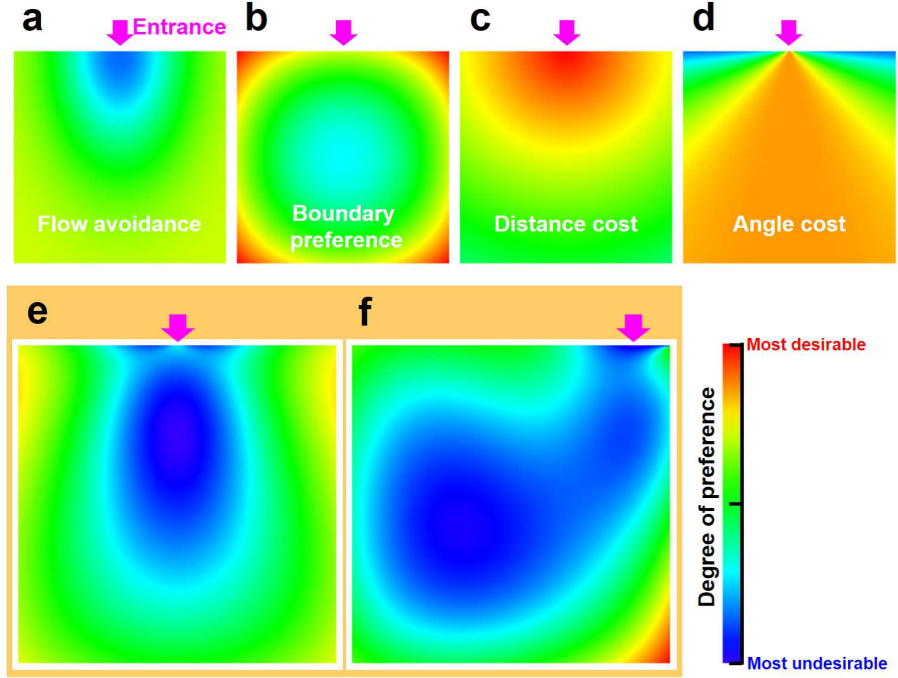}
\caption{Schematic representations of the factors explaining experimental results. (a) Flow avoidance (Eq. (\ref{fa})).
(b) Boundary preference (Eq. (\ref{bp})). (c) Distance cost (Eq. (\ref{dc})). (d) Angle cost (Eq. (\ref{ac})).
(e, f) Preference distribution obtained
  by superposing these four factors (a--d). For both panels (e, f),
  the identical parameter values were used, except for the entrance
  position. Parameters were set as $(f,f_1^2,f_2^2,b,d,a) =
  (0.6,0.1,0.3, 3, 0.4, 0.05)$ for a dimensionless area, $(x,y) \in
  [-0.5,0.5]\times [-1,0]$.}
\label{sch}
\end{figure}

Each factor is assumed as follows.
For comfortableness, pedestrians prefer positions where less disturbance is expected. 
Here, we consider avoidance factors against two types of interferences that are suggested by the occupation order shown in Fig.~\ref{figs} and Fig.~\ref{ir}b.
The first is (i) \textit{flow avoidance}, i.e., the desire to avoid any dynamical interference into the private space
caused by succeeding pedestrians passing by to reach their desired
positions. This factor should be salient near the entrance, as schematically shown in Fig.~\ref{sch}a. 

The other factor is psychological pressure in the (expected) final state. The degree of discomfort caused by the presence of other persons is known to strongly depend on the distance between them (i.e. the local density) \cite{Worchel1976}. 
In this measure, the boundaries, in particular corners,
are preferable because the level of discomfort is lower due to
  fewer neighbors (Fig.~\ref{sch}b; see also Appendix \ref{boundary preference}). Hence we refer to the second factor as
(ii) \textit{boundary preference}.  Similar behavior has been reported in other situations, e.g., seat preference in trains \cite{Evans2007} and classrooms \cite{Kaya2007}.

The presence of flow avoidance is supported by the fact that in scenario~C, the area in front of the entrance is less preferred,
even though it is near a boundary (Fig.~\ref{ir}b). Furthermore, the boundary preference is clearly seen in the same figure as the middle area, which is not directly affected by the inflow, is less preferred than the boundaries.

Additional contributions are (iii)
\textit{distance cost} and (iv) \textit{angle cost}.  The distance
cost takes into account the effort that has to be made to reach a
comfortable position. A position becomes more unattractive if it is farther
away from the entrance. It is natural to assume that this distance
factor is isotropic as shown in Fig.~\ref{sch}c.  The dependence on
the distance is less obvious and not necessarily linear. 
Distance cost explains why the area that is most distant from the entrance (bottom-left) is not preferred
in scenario~C, although it should be most attractive from the perspective of flow avoidance and boundary preference (Fig.~\ref{ir}b).

The \textit{angle cost} takes into account the effort necessary when
  changing the direction of motion. When a pedestrian selects a
position corresponding to a large angle $|\theta_i|$, he or she has to
immediately change the walking direction to directly reach there.
This leads to additional energy consumption and consequently discomfort. During
the experiments, in many cases the participants avoided immediate
turning but instead chose a longer but more comfortable path following an arc with a gradual angle variation.
In addition, the approach to such positions requires immediate
decision-making.  Any delay might increase the walking distance.  This
can be clearly observed in the experiment (see Fig.~\ref{ir}c). 
Therefore it is reasonable to believe that the
angle cost is especially salient for large angles, as shown in
Fig.~\ref{sch}d. This factor explains why the preference exhibits distinctive patterns 
in the areas near the two corners, i.e., the top-left and bottom-right corners in scenario C (Fig.~\ref{ir}b).
These two areas have the same conditions of distance from the entrance and boundary preference.
Also, the flow avoidance factor should not be effective in these two areas.

We suggest that these intuitively plausible factors (i)--(iv) predominantly
affect the pedestrians decision-making behavior. For a demonstration, we simply superposed these factors 
for scenarios N and C (Fig.~\ref{sch}e and Fig.~\ref{sch}f), choosing parameter values such that realistic patterns were qualitatively reproduced.
We want to emphasize that this interpretation by the four factors is at least capable of reproducing realistic patterns and might be useful
for understanding the underlying human collective behavior. It is not
designed to quantitatively predict each pedestrian's decision. Thus, the functions and parameters used in
Fig.~\ref{sch}a--Fig.~\ref{sch}d should be considered as rough qualitative estimates. 
Furthermore it needs to be critically checked whether these four factors are additive or not.  However, this
  formulation contributes rich insights to the understanding of the inflow process and serves as starting point for further
  investigations of this problem.

\section{Discussions}\label{discuss}
\begin{figure}[h]
\centering \includegraphics[width=140mm]{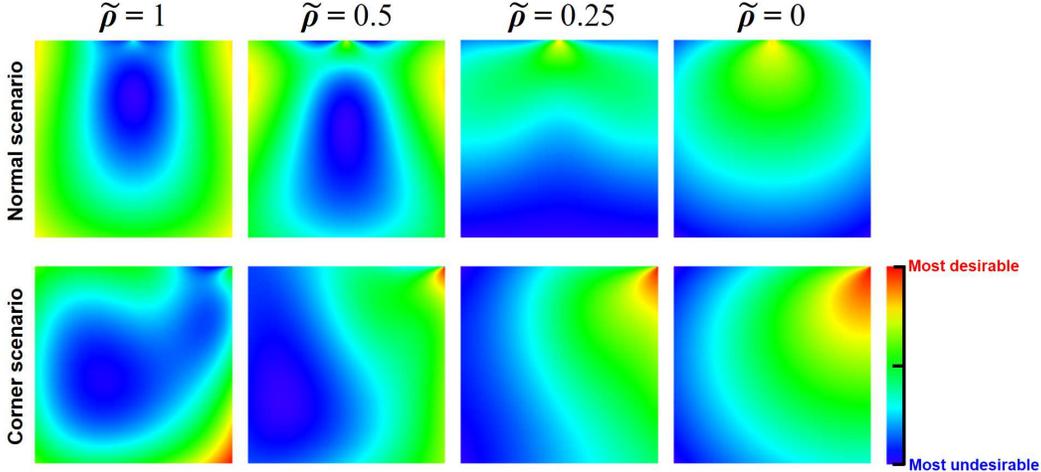}
\caption{Predicted location preference for various congestion levels, $\tilde{\rho}$.
  By assuming that the flow avoidance is proportional to the number of
  pedestrians, we set $f = \tilde{\rho} f_0$, where $f_0(=0.6)$ is the
  strength of the flow avoidance corresponding to $\tilde{\rho} =1$
  (25 pedestrians).  The other parameters were left unchanged:
  $(f_1^2,f_2^2,b,d,a) =(0.1,0.3,3,0.4,0.05)$.}
\label{pednum}
\end{figure}

\begin{figure}[h]
\centering
\includegraphics[width=140mm]{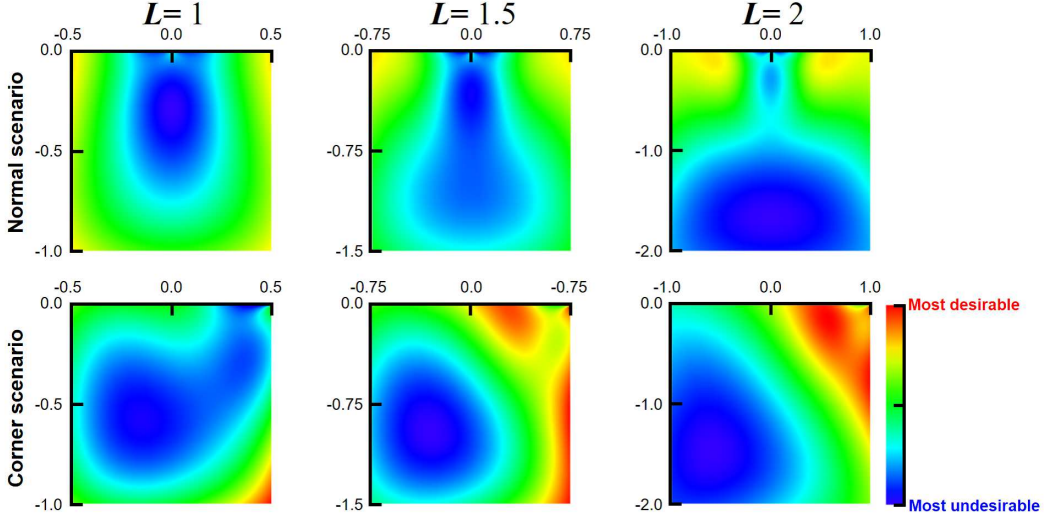}
\caption{Predicted location preference for 
  various area sizes, $\tilde{L}\times\tilde{L}$, with the same number
  of pedestrians. The condition $\tilde{L}=1$ corresponds to the size
  of the original experimental area in the main text, 3.6 m $\times$
  3.6 m.  According to the extension of the area, the normalized
  density is reduced as $\tilde{\rho} = \tilde{L}^{-2}.$ The other
  parameters were left unchanged: $(f, f_1^2,f_2^2,b,d,a) =(0.6,
  0.1,0.3,3,0.4,0.05)$.}
\label{areasize}
\end{figure}

We reported our experimental findings on the inflow process for three different scenarios, i.e., with two different entrance positions and with an obstacle. 
We observed that the position of the entrance had a significant impact on the pedestrians' choice of location, whereas the obstacle did not largely affect pedestrian movement.
In the experiments, we intentionally selected the number of
participants and the area size (i.e., the average final density) such that inter-pedestrian interactions
were of intermediate strength.
Although its validity should be further investigated, our description of the results using the four factors suggests
expected preference patterns in different scenarios, which could be considered as a base-line reference for future studies.  For
example, when only a small number of pedestrians enter the area (and
they know that), expected pressure will be small, which would
result in approximately no preference for boundaries.  Moreover, flow avoidance
would be less important, making positions near the entrance more attractive. 
This becomes obvious in the limiting case of a single passenger in an elevator (see Fig.~\ref{pednum} for example visualizations).
Meanwhile, when the area is very large, the
preference for the boundary would be outweighed by
the distance cost, and the area close to the entrance would be
preferred (see Fig.~\ref{areasize} for example visualizations).  
Furthermore, for people in cultures with high tolerance against interference of
personal space, the boundary preference factor might become weaker, which
should be investigated via cross-cultural experiments. 
Also, the presence of different types of individuals (e.g., females and children) and social groups (e.g., families and friends) \cite{Moussaid2010, Bode2015,Gorrini2015}
may affect their boundary preference.  
Therefore, such differences might be understood through the four factors, although our experimental results are based on limited scenarios.
In reality, other attractions are often observed. For example, in
trains and buses in rush hours, exit doors are more attractive for
passengers who attempt to leave first, which has been experimentally confirmed \cite{Liu2015,Liu2016}. Such behavioral differences may appear based on the destination of each passenger. 
In future studies, for which
our findings are useful for providing a base-line case, such a
trade-off should also be investigated.  Furthermore, the alignment of
head direction to an entrance could be diverted by
attractive objects. When there is a common understanding of the next
motion (e.g., leaving through a separate exit) passengers are attracted to that direction and align their heads accordingly. 

Detailed models of pedestrian motion inevitably require to take unobservable internal states of humans into account, e.g., for decision
  processes and interactions between pedestrians. 
Our findings suggest some principles in evaluating the current and future states around each individual.
Pedestrians try to avoid interference in their
personal space, considering the cost that depends on physical
constraints (distance, angle, etc.).  Such future anticipation cannot
be captured by one of the current paradigms of
pedestrian motion that is described by a driving force to a predefined
target and local avoidance \cite{Helbing2000,Yu2005,Chraibi2010}, but requires models that include
  decision processes, like most cellular automata \cite{Kirchner2002}.  For this reason, the inflow process has not received much attention in
pedestrian dynamics research until now. 

A theory of space evaluation of pedestrians
could provide a missing link for better understanding pedestrian
motion.  In addition, it could potentially lead to important
applications for designing comfortable and safe facilities.  In our
experiment, the shape of the area was suitable for attaining a
uniform pedestrian distribution.  Meanwhile, space in mass transport
systems (e.g. trains and buses) is sometimes not open due to the
presence of seats and hand rails that could distort or blind
pedestrians' space recognition, and lead to inefficient inhomogeneous
distributions and local overcrowding.  In our
  experiments, an additional obstacle had only a small effect on pedestrian trajectories, while leaving the
  system dynamics mostly unaffected. However, the final
  distribution and the location preference could change if the
obstacle were placed at a more obstructive location (e.g. just in front
of the entrance), or if several obstacles were placed in the area. This is a topic of future research.
\newpage

\appendix
\section*{Appendix: Materials and Methods}
\section{Trajectory detection}
To extract the trajectory of each individual person, the software
PeTrack was used \cite{Boltes2013}.  After deinterlacing the image,
the lens distortion was removed.  Then, the extrinsic calibration,
i.e., the camera position and angle of view according to the moving
plane, was performed for measuring the real position of each
head.  The perspective distortion had only a minor influence on the
calculated position because the viewing angle was small owing to the
high mounting position of the camera at a height of 21 m. Hence an average
body height of 1.73 m for all persons could be assumed.  The black and
red dots on the white cardboard worn on each head were detected by
searching for directed isolines of the same brightness with a
subsequent analysis of their shape (size, relative position to each
other, and aspect ratio).  The isovalue for the brightness threshold 
varied over a large interval to cope with the different
lighting conditions during the experiment.  The two dots were applied
to determine the orientation of the head, where the red dot indicated
the face side of the head (Fig.~\ref{set}d).  The position of each
person was defined to be in the middle of the black and red dots.  To
track a person over time, similar pixels related to the marker on
the head were searched in successive frames such that the detected
positions could be concatenated correctly over time for the resulting
trajectory.

\section{Time-averaged density profile}\label{density profile}
Fig.~\ref{ir}b was produced by averaging the temporal density of
pedestrians whose walking-velocity values were smaller than 0.15 m/sec, over entrance time and trials. In each frame, a set of
pedestrian positions in the experimental area defines the Voronoi
diagram (Fig.~\ref{figs}b; see also Ref. \cite{Zhang2011}). Let $A_i$ be
the area of each Voronoi cell for pedestrian $i$ (who is in the
experimental area). The density at a point ($x,y$) is defined as
\begin{equation}
\rho(x,y) =\frac{1}{A_i}\qquad {\rm if}\quad (x,y)\in A_i.\label{ld}
\end{equation}
First, we divided the experimental area into squares of side 0.2 m.  For each
square $j$, the time-averaged density ($\bar{\rho_j}'$) was calculated using
the Voronoi diagram for each trial.  In this procedure, we used the
trajectory data of eight frames per second between the entrance times of
the first and final pedestrians.  Then after normalizing the average
values to neutralize the difference in entrance time ($\bar{\rho_j}'
\rightarrow \bar{\rho_j}$ such that the sum over the experimental area
becomes a constant value $\sum_j0.2^2\bar{\rho_j} = \frac{25}{2}$ for
every trial
%\footnote{When the number of pedestrians in the area, $n$, increases at a constant rate from $n=0$ to $n=25$, the time average of $n$ is $\frac{25}{2}$.})
, they were averaged over trials.  If the
area is uniformly used and pedestrian inflow rate is constant, the
time averaged density is calculated as $\langle\bar{\rho_j}\rangle =
\frac{1}{2}\frac{25}{3.6^2}=0.965\;\rm{ped}/\rm{m}^2$.

\section{Deviation of walking trajectory from the shortest path}
We define $s_i$ as the walking distance of pedestrian $i$ from
the entrance to the final position.  If a
pedestrian directly moves to his final position, it coincides with the
shortest distance $r_i =|{\bm r}_i(t_{\rm{f}})|.$ Here ${\bm r}_i(t)$
is the position vector of pedestrian $i$ at time $t$, and $t_{\rm{f}}$
is the time when a pedestrian reached the final position, which was collected using the velocity threshold condition, $t_{\rm{f}} =
\min\{ t > t_{\rm{e}} ||\dot{{\bm r_i}}(t)| < 0.15\;\rm{m}/\rm{sec}\}$
in practice (Fig.~\ref{ir}c).  Here, $t_{\rm{e}}$ represents the
pedestrian's entrance time.  Using these variables, the degree of the 
walking path deviation is defined as
\begin{equation}
\frac{s_i}{r_i} = \frac{\int_{t_{\rm{e}}}^{t_{\rm{f}}} 
|\dot{\bm r}_i (t)| dt}{|{\bm r}_i(t_{\rm{f}})|},\label{nw}
\end{equation} 
which is, by definition, always larger than $1$.

\section{Schematic representations of location preference}
To capture the elusive features of pedestrian interactions, we attempt
to illustrate the factors that determine the choice of locations, and
thereby reproduce Fig.~\ref{ir}b by considering a simple superposition
\begin{equation}
P(x, y) =  F +B+ D +A,\label{pref}
\end{equation}
where $P, F,B,D$ and $A$ are abstract cost functions for the
location preference, the gain by the flow avoidance, the boundary
preference, the distance cost and the angle cost, respectively.  
These factors are defined to increase as
the location becomes more desirable for pedestrians.  Of course, these
factors are not necessarily additive, but it is natural to consider
that $P$ is a monotonically increasing function of $F,W,D$ and $A$.
As the simplest form, we tentatively adopt this definition of $P$ in
this article.

\subsection{Flow avoidance}
To express the flow avoidance we assume a Gaussian function (Fig.~\ref{sch}a)
\begin{equation}
F(x,y) = -f \exp{\left(-\frac{(x-x_0)^2}{f_1^2} - 
              \frac{(y-y_0)^2}{f_2^2}\right)},
\label{fa}
\end{equation} 
where $f$, $(x_0, y_0)$, and $(f_1,f_2)$ represent a positive
constant, the location of the entrance and the widths of flow
avoidance, respectively. Considering the direction of flow, it is
natural to presume $f_1 < f_2$. Pedestrians might anticipate other
pedestrians' choice, which would affect this flow avoidance. However,
it is a formidable challenge to include such feedback.  Thus we here
forbear to go into detail and instead define $F$ a priori.  Since this
factor only reduces the preference near the entrance, the conclusion
in this study is not significantly affected by a specific choice of
this function. 

\subsection{Boundary preference}\label{boundary preference}
When describing the psychological pressure between pedestrians, it is natural to consider a function $g$ that
decays with the distance between two pedestrians (Fig.~\ref{dc}). 
In the uniform pedestrian distribution preconceived by a pedestrian, the expected number of pedestrians
in an area $dA$ at position $\bm{r'}\in D$ ($D:$ experimental area) is denoted by $\tilde{\rho}dA$, 
where $\tilde{\rho}$ is the normalized density (pedestrians per area). 
This area provides pressure against position $\bm{r}$, whose magnitude is $g(|{\bm r}-{\bm r'}|)\tilde{\rho} dA.$
Thus, the total (expected) psychological pressure on position $\bm{r}$ is obtained by taking the sum of this factor for 
all the positions ($\bm{r'}$) in the experimental area. 
For simplicity we take the limit of $dA\rightarrow 0$, which yields
\begin{equation}
B({\bm r}) = -\int\int_{\bm{r'}\in D} g(|{\bm r}-{\bm r'}|)\tilde{\rho} dA\label{bp}.
\end{equation}
Because the area out of the boundaries does not contribute to this value, 
the areas of boundaries become more preferable (see Fig.~\ref{sch}a). 
Thus, this factor explains pedestrians' preference to the boundaries.
The pedestrian density is normalized as $\tilde{\rho} = 1$ in this study (for 25
pedestrians in a 3.6~m$\times$ 3.6~m area).  
Here, we tentatively
defined the discomfort function as $g(x) =b (1+x)^{-2}$ (with a positive
constant $b$), avoiding a singularity at $x=0$ because pedestrians do not
normally fear being physically overlapped (at the densities considered
here).  It could also be defined
with the singularity, for which case, the integral should be performed
except for human's body area $|{\bm r}-{\bm r'}| < r_0$ (body radius).
Both definitions produce qualitatively the same preference, and thus
we adopted the simpler one because our interest is not in finding an
exact description of the factor.

\begin{figure}[tbhp]
\centering
\includegraphics[width=75mm]{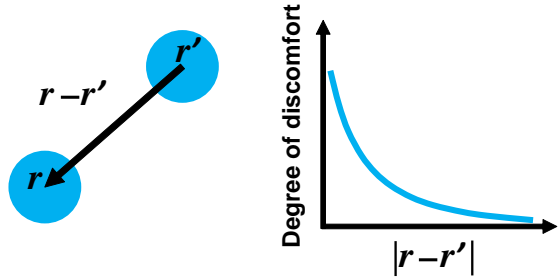}
\caption{Closeness and degree of discomfort. In this article we assumed 
  the following function: $g(x) =b (1+x)^{-2}$, where $x$ represents
  the distance between the two individuals (see also Eq. (\ref{bp})).}
\label{dc}
\end{figure} 

\subsection{Distance cost}
The distance cost should be a monotonically increasing function of $|\bm{r}|.$
For simplicity we assume the linear expression
\begin{equation}
  D({\bm r}) = - d |{\bm r}|,\label{dc}
\end{equation} 
with a positive constant $d$ (Fig.~\ref{sch}c). 
 
\subsection{Angle cost}
For the angle cost we selected a function that nonlinearly decreases
 with the absolute value of $\theta$:
(Fig.~\ref{sch}d)
\begin{equation}
A({\bm r}) = -a|\theta|^\alpha.\label{ac}
\end{equation} 
Here $a$ and $\alpha$ are positive constants. In this article, we
tentatively assume $\alpha=3$.

We emphasize that the absolute values of $F,B,D$, and $A$ are not
essential, but their relative strengths are important. 
As the exact forms of these functions cannot be easily determined, we used simple functions that satisfy the requirements discussed in Sec. \ref{interpretations}. 
In
Fig.~\ref{sch}, we selected the parameter values as $(f,f_1^2,f_2^2,b,d,a) =
(0.6,0.1,0.3, 3, 0.4, 0.05)$ for a dimensionless area, $(x,y) \in
[-0.5,0.5]\times [-1,0]$. 
These parameter values were manually set such that the density patterns become consistent with the experimental results, as a rough estimate.
 Examples with different parameter values
are shown in Fig.~\ref{pednum} and Fig.~\ref{areasize}.

\section{Ethics statement}
This experiment has been approved by the ethics committee of the office for life science research ethics and safety, 
The University of Tokyo.
All participants provided written informed consent to participate in
the experiment.

\begin{acknowledgements}
  We acknowledge Haruki Ishikawa for assistance in preparatory works
  on pedestrian detection.  We would like to thank Ryosuke Nishi and Marina Dolfin
  for their valuable
  comments on this manuscript.  TE, DY, and KN acknowledge financial
  support by JSPS Grant Number 13J05086, 15K17583 and 25287026,
  respectively.
\end{acknowledgements}

% BibTeX users use
%\bibliographystyle{apsrev} % mathematics and physical sciences
%\bibliography{inflow3} % name your BibTeX data base

% Non-BibTeX users please use
%\begin{thebibliography}{1}
%\bibitem{RefJ}
%  % Format for Journal Reference
%  Author: Title. Journal \textbf{Volume}, Pages (Year), \doi{XX.XXXX/XXX}
%\bibitem{RefP}
%  % Format for Proceedings Reference
%  Author: Title. In: Conference, Pages (Year), \doi{XX.XXXX/XXX}
%\bibitem{RefB}
%  % Format for Books
%  Author: Book title. Publisher (Year), \doi{XX.XXXX/XXX}
%\end{thebibliography}

\end{document}